\documentclass[doublecol]{epl2} 

\usepackage{amssymb}
\usepackage{amsfonts}
\usepackage{amsmath}
\usepackage{amsthm}
\usepackage{bbm}

\title{Hyperbolicity impedes emergence of chimera states in networks of nonlocally coupled chaotic oscillators}
\shorttitle{Hyperbolicity impedes emergence of chimera states} 

\author{N.~Semenova\inst{1} \and A.~Zakharova\inst{2} \and E.~Sch{\"o}ll\inst{2} \and V.~Anishchenko\inst{1}}
\shortauthor{N.~Semenova \etal}

\institute{                    
  \inst{1} Institut f{\"u}r Theoretische Physik, Technische Universit\"at Berlin, Hardenbergstra\ss{}e 36, 10623 Berlin, Germany\\
  \inst{2} Department of Physics, Saratov State University, Astrakhanskaya str. 83, 410012 Saratov, Russia\\
}

\abstract{We analyze nonlocally coupled networks of identical chaotic oscillators with either time-discrete or time-continuous dynamics (Henon map, Lozi map, Lorenz system). We hypothesize that chimera states, in which spatial domains of coherent (synchronous) and incoherent (desynchronized) dynamics coexist, can be obtained only in networks of nonhyperbolic chaotic systems and cannot be found in networks of hyperbolic systems. This hypothesis is supported by analytical results and numerical simulations for hyperbolic and nonhyperbolic cases.}

\begin{document}

\maketitle

\section{Introduction}
Chimera states in dynamical networks of nonlocally coupled chaotic oscillators have recently attracted
much attention \cite{PAN15}. They represent an intriguing phenomenon where an ensemble of identical elements
with symmetric coupling spontaneously splits into spatially separated coexisting domains of coherent
(synchronized) and incoherent (desynchronized) dynamics.  
Since their first discovery in systems of coupled phase oscillators \cite{KUR02a,ABR04}
they have been found in a broad range of diverse models \cite{MOT10,PAN15}, ranging from
time-discrete maps~\cite{OME11} and time-continuous chaotic models~\cite{OME12}, Stuart-Landau oscillators \cite{SET13,SET14,ZAK14,ZAK15b}, globally coupled lasers \cite{BOE15}, Van der Pol oscillators \cite{OME15a,BAS15} FitzHugh-Nagumo neural systems \cite{OME13, OME15}, population dynamical models \cite{HIZ15} to autonomous Boolean networks \cite{ROS14a}. 
They have recently also been observed experimentally, e.g., in 
optical light modulators \cite{HAG12}, chemical \cite{TIN12}, mechanical \cite{MAR13,KAP14}, electronic \cite{LAR13,GAM14},
optoelectronic\cite{LAR15}, and electrochemical \cite{WIC13,SCH14a}
oscillator systems. In real-world systems chimera states might play a role, e.g., in the unihemispheric sleep of birds and
dolphins~\cite{RAT00}, in neuronal bump states~\cite{LAI01,SAK06a}, in epileptic seizure \cite{ROT14}, in power
grids~\cite{MOT13a}, or in social systems~\cite{GON14}. 

In spite of intense theoretical study, no universal mechanism for the formation of chimera states in different systems has yet been
established. It remains unclear what the common features of the different specific models with different local dynamics are.
Chimera states emerge as a hybrid state on the transition between completely synchronized coherent and completely desynchronized incoherent states with decreasing coupling strength, and such coherence-incoherence bifurcation scenarios have been found to display unversal features in many different time-discrete and time-continuous models \cite{OME11,OME12,HAG12,OME15a}.

In this work we analyze networks of identical chaotic oscillators with nonlocal coupling, with the aim to understand the mechanism of the emergence of chimera states in different systems. We study networks of both time-discrete chaotic maps (Henon and Lozi map) and time-continuous three-variable chaotic systems (Lorenz model). In the present work we observe that the emergence of chimera states appears to be linked with the property of nonhyperbolicity of the dynamical system, while hyperbolic systems\footnote{These systems are reputed hyperbolic because of non-existence of stable regimes and its bifurcations.} appear not to exhibit chimeras. In a number of papers chimeras have been reported in ring networks of chaotic nonhyperbolic systems exhibiting a period-doubling route to chaos, e.g., the logistic map \cite{OME11}, the cosine map \cite{HAG12} and the R\"{o}ssler attractor \cite{OME12}, while chimeras of oscillating states have not been observed for the Lorenz attractor \cite{DZI13}, which is hyperbolic for the standard parameters.  
In particular we study the Henon map \cite{HEN76}, which is an example of a nonhyperbolic dynamical system, like the R\"{o}ssler attractor, and the Lozi map as an example of a hyperbolic system. By tuning the parameters of the Lorenz model we show that chimeras
arise in the nonhyperbolic regime, while they are suppressed in the hyperbolic regime.

\section{Hyperbolic and nonhyperbolic chaotic systems} 

Fist, as an example of a nonhyperbolic dynamical system, we study the H\'{e}non map \cite{HEN76}:
\begin{equation} \label{eq:henon}
 x^{t+1} = 1-\alpha (x^t)^2 +y^t, \qquad y^{t+1} = \beta x^t,
\end{equation}
where $t$ is the dicrete time, $x^t$, $y^t$ are the dynamical variables, and $\alpha>0$ and $\beta>0$ are parameters of the map. It should be noted that the case of $0<\beta <1$ corresponds to a dissipative diffeomorphism. 
The Henon map under high-ratio compression ($\beta \rightarrow 0$) reduces to the logistic map $x^{t+1}=1-\alpha (x^t)^2$. 
It belongs to the wide class of the Feigenbaum maps, characterized by a quadratic maximum. When the parameters are changed, the map (\ref{eq:henon}) demonstrates a period-doubling bifurcation cascade and periodic orbits which are in agreement with Sharkovsky ordering \cite{Sharkovskii}. 

\begin{figure}[h!]
\centering
\center{\includegraphics[width=1\linewidth]{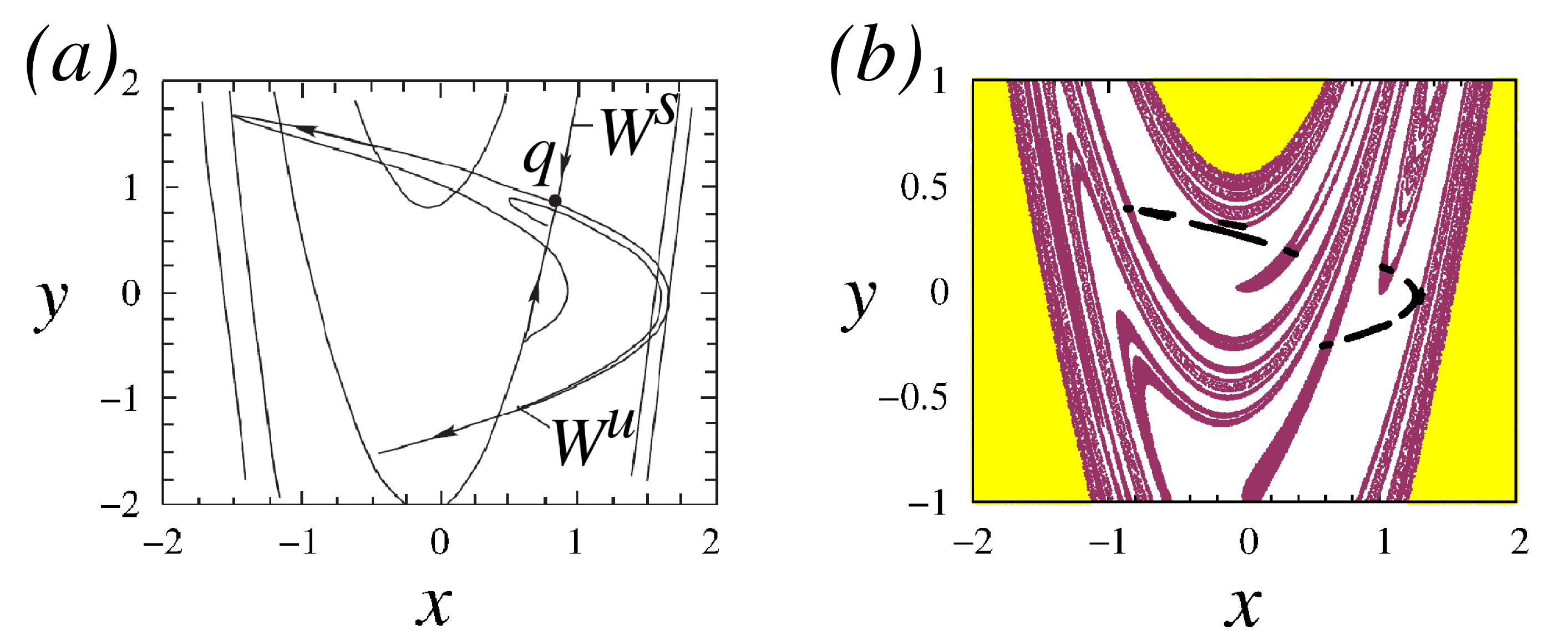}}
\caption[]{(a) -- Stable ($W^s$) and unstable ($W^u$) manifolds of a saddle point $q$ of the Henon map for $\alpha=1.3$ and 
$\beta=0.3$. \\ (b) -- Attractors in the Henon map and their basins of attraction in the phase plane $(x,y)$ 
 for $\alpha=1.078$ and $\beta=0.3$. The white region is the basin 
of attraction of the four-band chaotic attractor (black dots), the purple (dark grey) region corresponds to the basin 
of attraction of the six-band chaotic attractor, and trajectories from  the yellow (light grey) regions diverge to infinity.} 
\label{fig1ab}
\end{figure}

The Henon map can be used for a qualitative description of bifurcation phenomena in three-dimensional time-continuous systems of 
spiral-type chaos because of the saddle-focus separatrix loop. As shown, for example, in \cite{ANI95} the spiral-type chaos in a three-dimensional system gives a two-dimensional map in the Poincare cross section, equivalent to Eq.(\ref{eq:henon}). The Henon map (\ref{eq:henon}) is an example of a nonhyperbolic dynamical system with bifurcation of homoclinic connections of stable and unstable saddle point separatrices \cite{ANI14}. This bifurcation is the basis for the emergence of the nonhyperbolic attractor. It means that the Henon map is the simplest model of a chaotic map which can be obtained as the Poincare section of the nonhyperbolic class of R\"{o}ssler oscillators. Figure~\ref{fig1ab}(a) illustrates the effect of homoclinic tangency of the stable ($W^s$) and unstable ($W^u$) saddle point manifolds in the Henon map. The homoclinic tangency leads to the complex structure of the Henon attractor shown in Fig.~\ref{fig1ab}(b). The nonhyperbolic attractor of the Henon system is characterized by the coexistence of several regular and chaotic attracting sets. Their attractor basins (white and purple regions) have a fractal structure. Upon change of parameters, these sets undergo various bifurcations, which are accompanied by drastic changes in the structure of the attracting set.

\begin{figure}[h!]
\centering
\center{\includegraphics[width=1\linewidth]{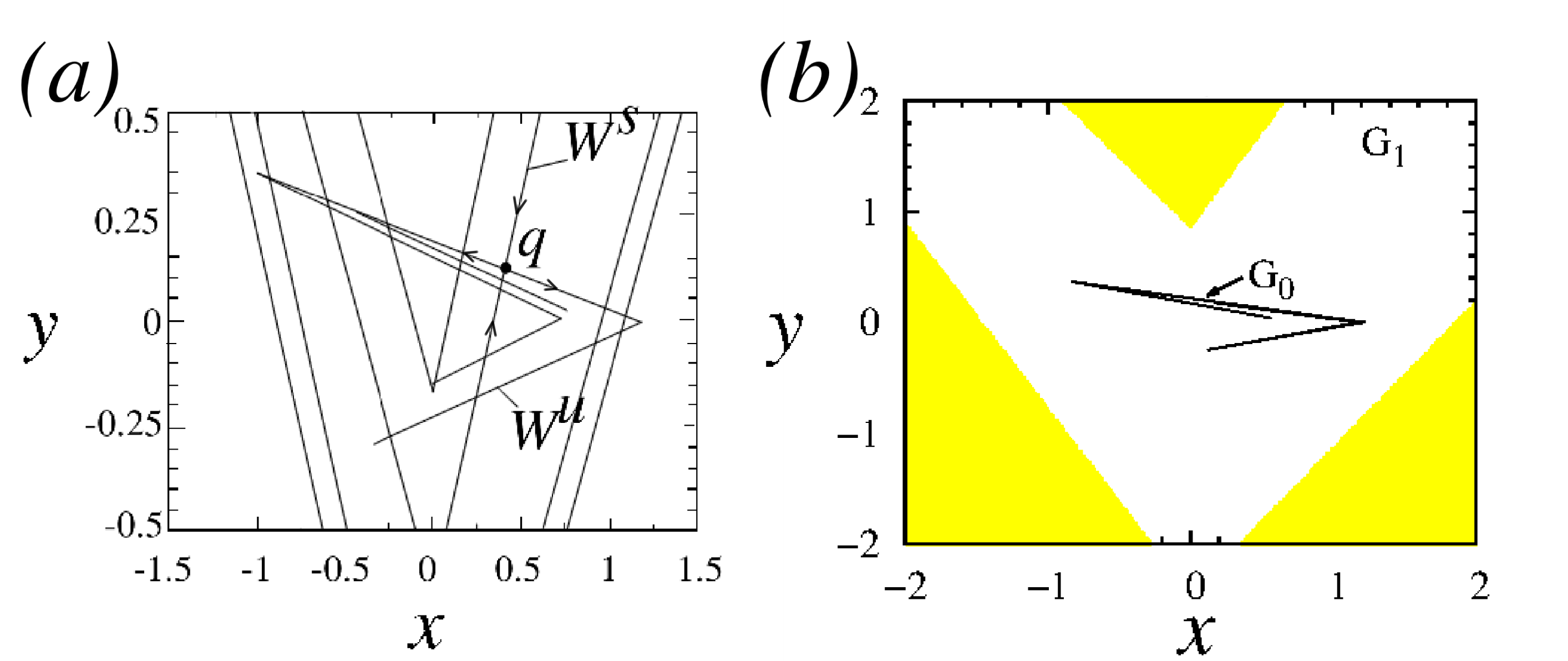}}
\caption[]{(a) -- Stable ($W^s$) and unstable ($W^u$) manifolds of a saddle point $q$ in the Lozi map for $\alpha=1.7$ and
$\beta=0.3$. \\ (b) -- Lozi attractor $G_{0}$ and the basin of its attraction $G_{1}$ (white) for $\alpha=1.5$ and $\beta=0.3$. 
Trajectories from the yellow (light gray) region diverge to infinity.} 
\label{fig2ab}
\end{figure}

The time-continuous Lorenz system belongs to another class of dynamical systems. It is an example of a hyperbolic attractor without regular stable subsets like equilibrium states or limit cycles. The Lorenz attractor is resilient to slight changes in parameters and perturbations of the system equations. The Lozi map \cite{Lozi} belongs to the class of chaotic maps which can be obtained as the
Poincare section of the hyperbolic class of Lorenz-type attractors:
\begin{equation} \label{eq:lozi}
x^{t+1}= 1-\alpha |x^t| +y^t, \qquad y^{t+1}=\beta x^t.
\end{equation}
The stable and unstable saddle point manifolds of the Lozi attractor intersect transversally (without tangency). The property of transversality is resilient to parameter variations. This means that the hyperbolic Lozi attractor is robust (structurally stable). It is illustrated in Fig.~\ref{fig2ab}(a). 

Figure~\ref{fig2ab}(b) shows that the Lozi attractor is a unique connected attracting set with bounded and homogeneous attractor basin in the phase plane. The attractor topology is robust to slight changes in the control parameters.


\section{Dynamics of ring networks of nonlocally coupled Henon and Lozi maps}
We consider a ring of $N$ nonlocally coupled two-dimensional maps, described by:
\begin{equation}\label{eq:coupling}
\begin{split}
x^{t+1}_i&=f(x^t_i,y^t_i)+\frac{\sigma}{2P}\sum\limits_{j=i-P}^{i+P} [f(x^t_j,y^t_j)-f(x^t_i,y^t_i)], \\
y^{t+1}_i&=\beta x^t_i,
\end{split}
\end{equation}
where $t$ is the discrete time, $N$ is the number of elements in the ring, $i=1,2\dots N$ is the index of the element, all indices are modulo $N$, $\sigma$ is the coupling strength, $P$ is the number of neighbours on either side, and 
$f(x,y)=1-\alpha x^2+y$ (Henon map, Eq.(\ref{eq:henon})), or $f(x,y)=1-\alpha |x|+y$ (Lozi map, Eq.(\ref{eq:lozi})).
We introduce the parameter $r=P/N$ as the coupling radius.

\begin{figure}[h!]
\centering
\includegraphics[width=1\linewidth]{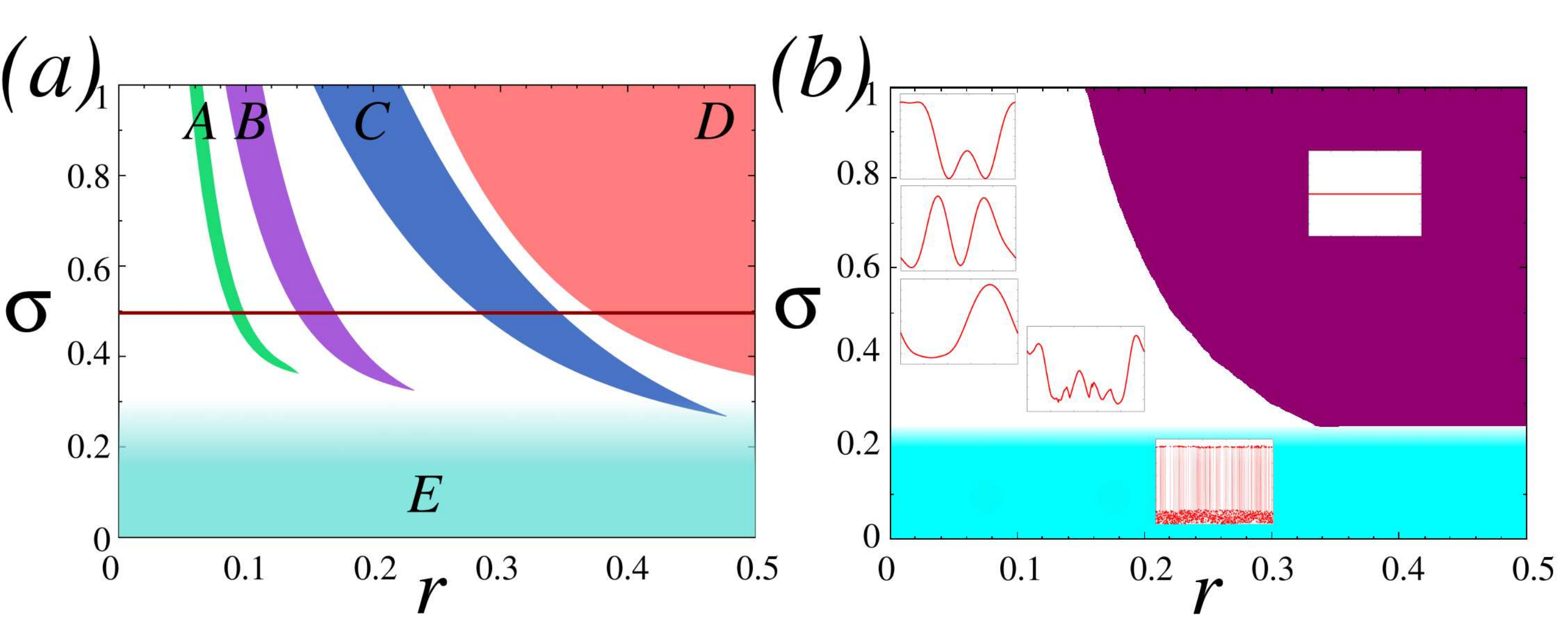}
\caption[]{Main regimes of nonlocally coupled Henon maps (a) and Lozi maps (b) in the $(r,\sigma)$ parameter plane. (a) -- Regimes of coherence with wave numbers $k=1,2,3$ (regions $C$, $B$, $A$, respectively). The horizontal line at $\sigma_c\approx 0.5$ marks the coherence-incoherence transition. (b) -- White region corresponds to multistable regime of stationary patterns and travelling wave profiles (insets show profiles with different wavenumbers for the same parameters and different initial conditions). Both systems can demonstrate completely synchronized chaotic state ((a),$D$ and (b),purple (dark) region) complete spatial incoherence ((a),$E$ and (b),blue (light) area). Parameters: $\alpha=1.4$, $\beta=0.3$, $N=1000$.}
\label{fig3}
\end{figure}

First, let us analyze the nonlocally coupled Henon maps (\ref{eq:henon}). Figure~\ref{fig3}a illustrates all main regimes of this system. It shows that the system of coupled Henon maps demonstrates the same regimes as coupled logistic maps 
\cite{OME11}. Their $(r,\sigma)$ parameter planes have the same qualitative and quantitative structure with coherence tongues ($A,B,C$).
The area $E$ in Fig.~\ref{fig3}a corresponds to the regime of the spatially incoherent states, $D$ is the region of the completely synchronized chaotic states. The regions $A$, $B$, and $C$ correspond to the spatially coherent states with wave numbers $k=3$, $k=2$, and $k=1$, respectively.

\begin{figure}[htbp]
\center{\includegraphics[width=.7\linewidth]{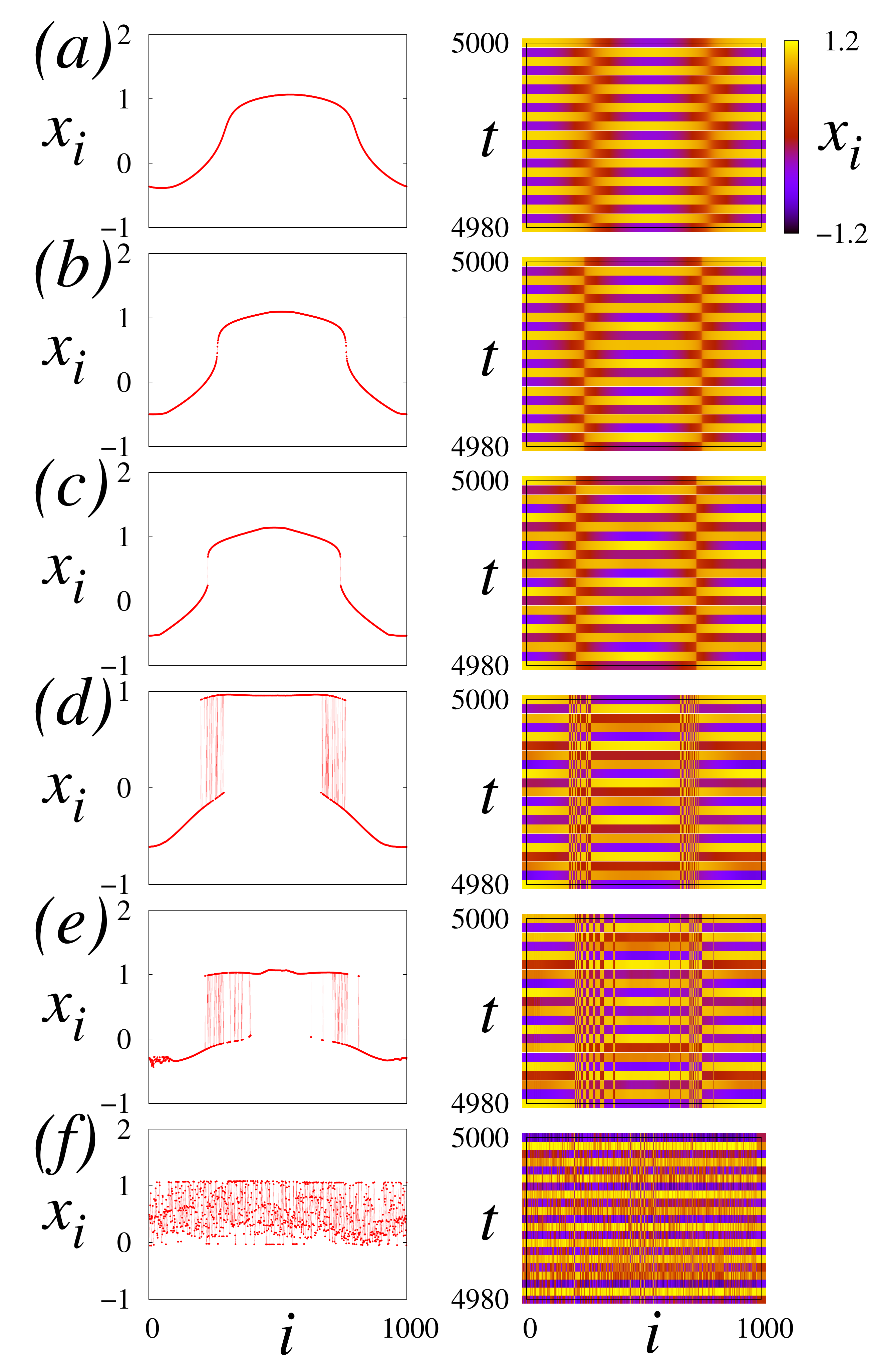}}
\caption{Coherence-incoherence bifurcation for coupled Henon maps for coupling radius $r=0.3$ and decreasing coupling strength 
(a) $\sigma=0.53$, (b) $\sigma=0.47$, (c) $\sigma=0.45$, (d) $\sigma=0.32$, (e) $\sigma=0.29$, (f) $\sigma=0.2$). Snapshots at time $t=5000$ (left columns) and space-time plots (right columns) are shown. Other parameters are as in Fig.~\ref{fig3}a.}
\label{fig4}
\end{figure}

By an analytical argument similar to \cite{OME12,HAG12} (Appendix) one can calculate the critical value of the coupling strength $\sigma_c$ (horizontal line in Fig.~\ref{fig3}a, below which the spatial wave profile ($k=1$, period-2 in time) in the coherence tongues becomes discontinuous, and a chimera state appears:
\begin{equation}\label{eq:sigma_crit_henon_main}
\sigma_c = 1 - 1/|f_x (x^* ,y^*)-\beta| = 1 - 1/|2 \alpha x^*+\beta|.
\end{equation} 
where $f_x$ is the derivative of $f$ with respect to $x$, and ($x^*, y^*$) are fixed points of the map.
The analytical result $\sigma_c \approx 0.52$ agrees very well with the numerically obtained $\sigma_c$ in Fig.~\ref{fig3}a
and with the snapshots (left column) and the space-time plots (right column) shown in Fig.~\ref{fig4} for the transition from coherence (Fig.~\ref{fig4}a) to incoherence with decreasing coupling strength.  In the tongue $C$ of Fig.~\ref{fig3}a for $\sigma<\sigma_c$
 two spatially incoherent domains appear (Fig.~\ref{fig4}b) and become wider (Fig.~\ref{fig4}c-e) with decreasing coupling strength $\sigma$ and, eventually, the dynamics becomes completely incoherent (Fig.~\ref{fig4}f). This scenario represents a ``coherence-incoherence'' transition. Similar results were obtained in \cite{OME11} for ring networks of logistic maps and R\"{o}ssler systems.

Figure~\ref{fig3}b illustrates the main regimes of nonlocally coupled Lozi maps in the $(r,\sigma)$ parameter plane. It shows a completely synchronized chaotic state (Fig.~\ref{fig3}b, purple (dark) region), a completely incoherent state (Fig.~\ref{fig3}b, blue (light) region) and regimes of stationary patterns and travelling spatial profiles (Fig.~\ref{fig3}b, white region). Similar results including travelling waves were obtained for the ring of coupled Lorenz systems \cite{DZI13}. It should be noted that chimera states have not been obtained in the ring of Lozi maps.
The critical value of the coupling strength for coupled Lozi maps, calculated by a similar argument as above under the hypothesis of spatial wavenumber $k=1$ and period-2 in time (see Appendix), is for $x^*<0$
\begin{equation}\label{eq:sigma_crit_lozi_main}
\sigma_c = 1 - 1/|\alpha -\beta|.
\end{equation} 
which yields $\sigma_c\approx 0.09$. However, this value is so small that the coupling cannot induce a coherent spatial profile with regular time-period-2 dynamics. Therefore the transition to partial spatial coherence does not exist.

\begin{figure}[htbp]
\center{\includegraphics[width=.7\linewidth]{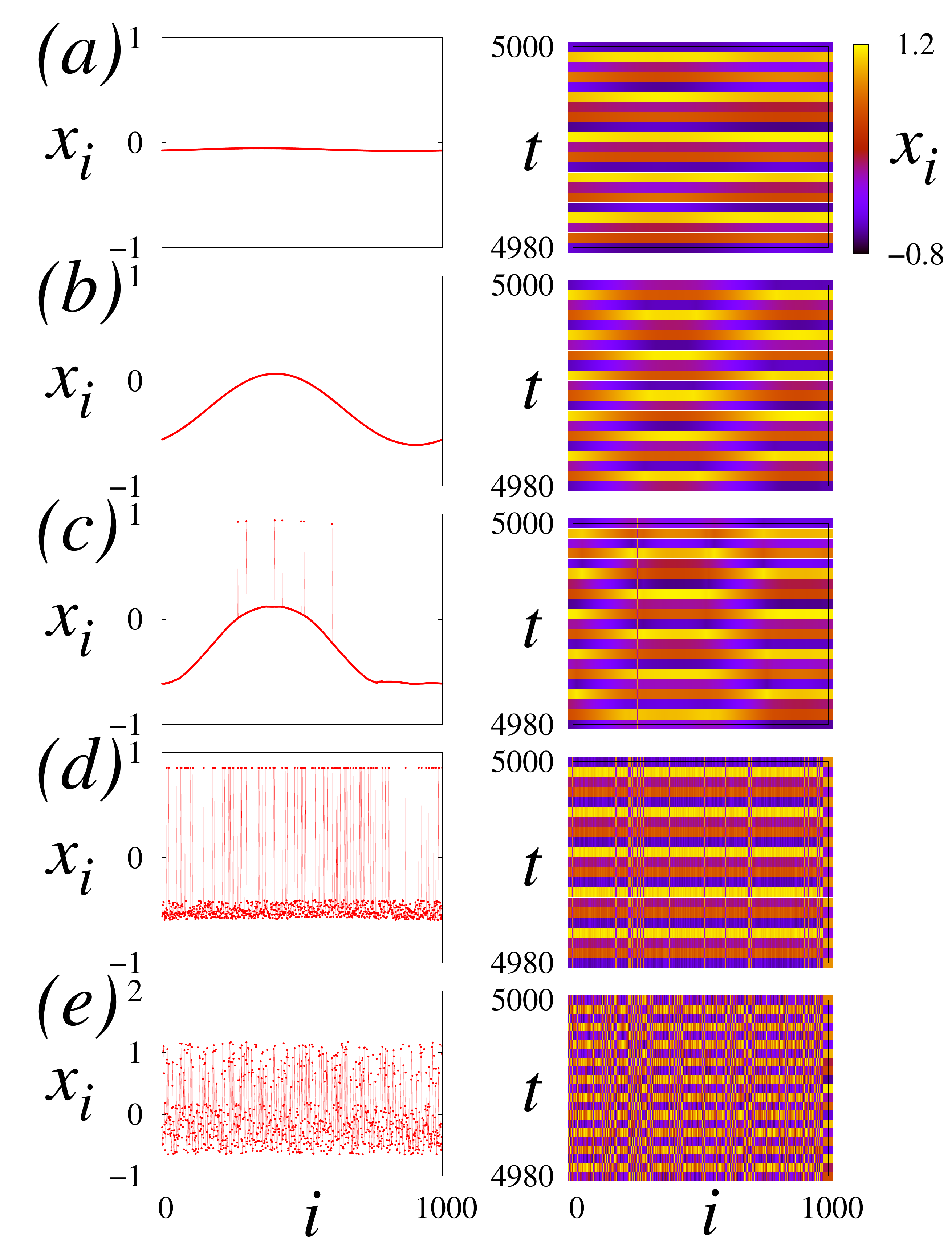}}
\caption{Coherence-incoherence transition for coupled Lozi maps for coupling radius $r=0.147$ and decreasing coupling strength 
(a) $\sigma=0.94$, (b) $\sigma=0.44$, (c) $\sigma=0.22$, (d) $\sigma=0.14$, (e) $\sigma=0.001$).
Snapshots at time $t=5000$ (left columns) and space-time plots (right columns) 
are shown. Other parameters are as in Fig.~\ref{fig3}b. Initial conditions were randomly distributed in square $x_{0i}\in(-0.5;0.5)$, 
$y_{0i}\in(-0.5;0.5)$.
}
\label{fig5}
\end{figure}

\begin{figure}[htbp]
\center{\includegraphics[width=.7\linewidth]{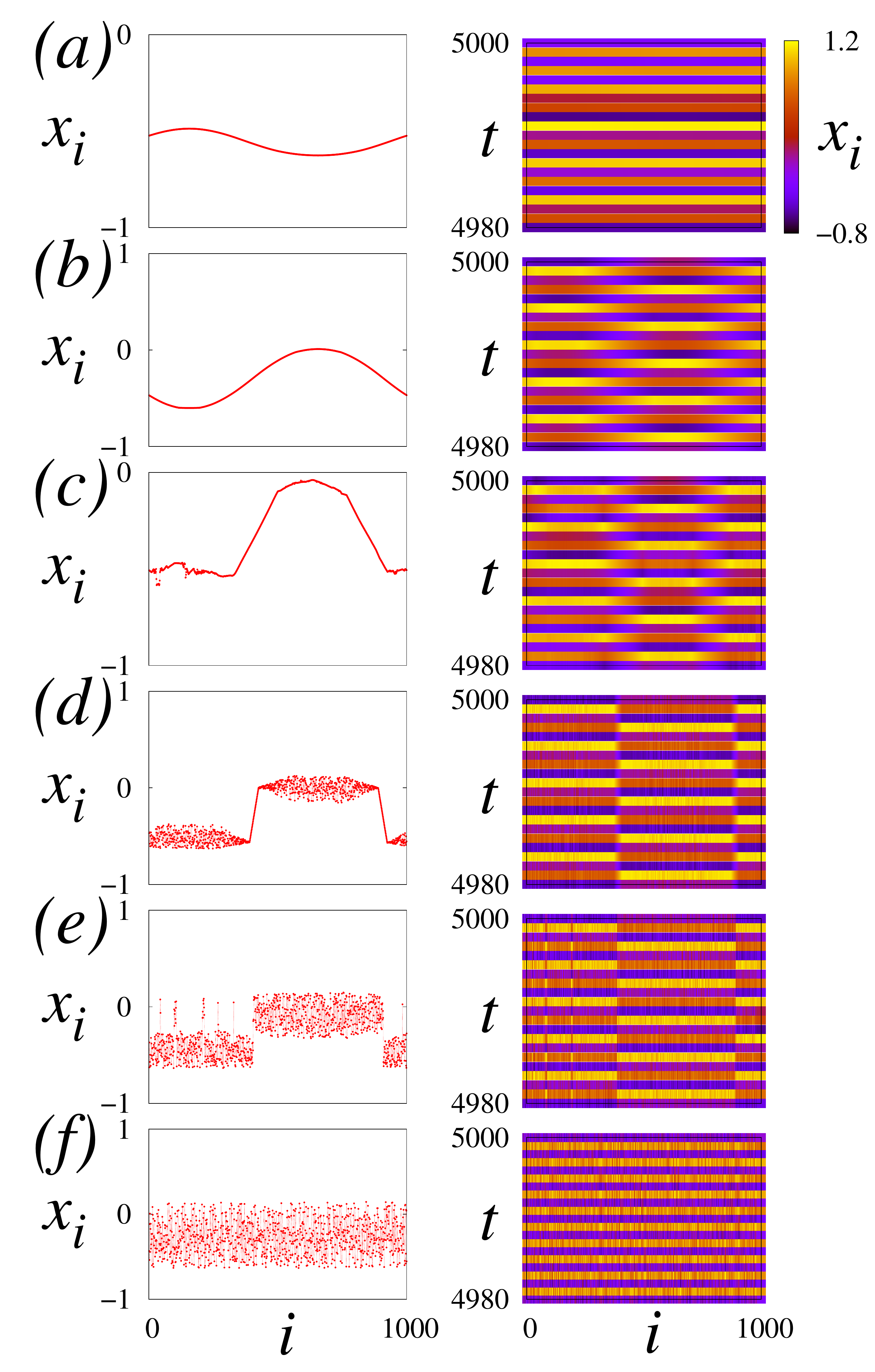}}
\caption{Coherence-incoherence transition for coupled Lozi maps for coupling radius $r=0.147$ and decreasing coupling strength 
(a) $\sigma=0.92$, (b) $\sigma=0.46$, (c) $\sigma=0.18$, (d) $\sigma=0.10$, (e) $\sigma=0.03$, (f) $\sigma=0.02$. Snapshots at time $t=5000$ (left columns) and space-time plots (right columns) are shown. Other parameters are as in Fig.~\ref{fig3}b. Initial conditions were prepared as $k=1$ spatial wave profile.  
}
\label{fig6}
\end{figure}

A more detailed analysis of the snapshots and space-time scenarios of coupled Lozi maps as protoptype of a hyperbolic system reveals completely different scenarios from complete coherence to complete incoherence with decreasing coupling strength, as shown in Fig.~\ref{fig5} for the case of random initial conditions and in Fig.~\ref{fig6} for specially prepared initial conditions corresponding to a $k=1$ spatial wave profile. In Fig.~\ref{fig5}a-b the profiles are coherent (completely synchronized), in
Fig.~\ref{fig5}c some solitary oscillators desynchronize, whose density increases (Fig.~\ref{fig5}d), until in  Fig.~\ref{fig5}e the
system is completely incoherent (desynchronized). In Fig.~\ref{fig6}a-c the profiles are again coherent, but Fig.~\ref{fig6}d,e  exhibits discontinuous spatial profiles, however, the incoherent parts are not confined to two incoherent domains located between two coherent domains as in nonhyperbolic systems, i.e., no chimera. Rather the whole spatial profile is incoherent, but the disorder is spread around the upper and lower branches.
This qualitatively different coherence-incoherence bifurcation occurs at a critical coupling strength which corresponds approximately to
the value found analytically in Eq.(\ref{eq:sigma_crit_lozi_main}). Finally, in Fig.~\ref{fig6}f complete incoherence is reached. 

\section{Coherence-incoherence transition and chimera states in nonlocally coupled Lorenz systems}
The absence of chimera states in a ring of coupled Lozi maps allows us to advance the following hypothesis. The chimera state cannot be obtained in a network of hyperbolic systems (both discrete and time-continuous). The Lorenz system is hyperbolic for standard parameters. Indeed, a detailed examination of a ring with nonlocally coupled Lorenz systems \cite{DZI13} confirms this hypothesis. This network exhibits complete synchronization, completely incoherent states, and stationary patterns or coherent travelling waves but no chimera states. All these regimes were also obtained in the network of Lozi maps, 
but we have not found chimera states. 

To test our hypothesis, we have examined the Lorenz system
\begin{equation} \label{eq:lorenz}
\dot{x}=-\gamma(x-y) \qquad \dot{y}=-xz+\rho x-y \qquad \dot{z}=-bz+xy,
\end{equation}
where $x, y, z$ are the dynamic variables, and $\gamma, b, \rho$ are parameters, in a range of parameters where the Lorenz system is nonhyperbolic, i.e., it has a spiral quasiattractor \cite{Bykov,ANI14}. 

\begin{figure}[htbp]
\centering
\includegraphics[width=.6\linewidth]{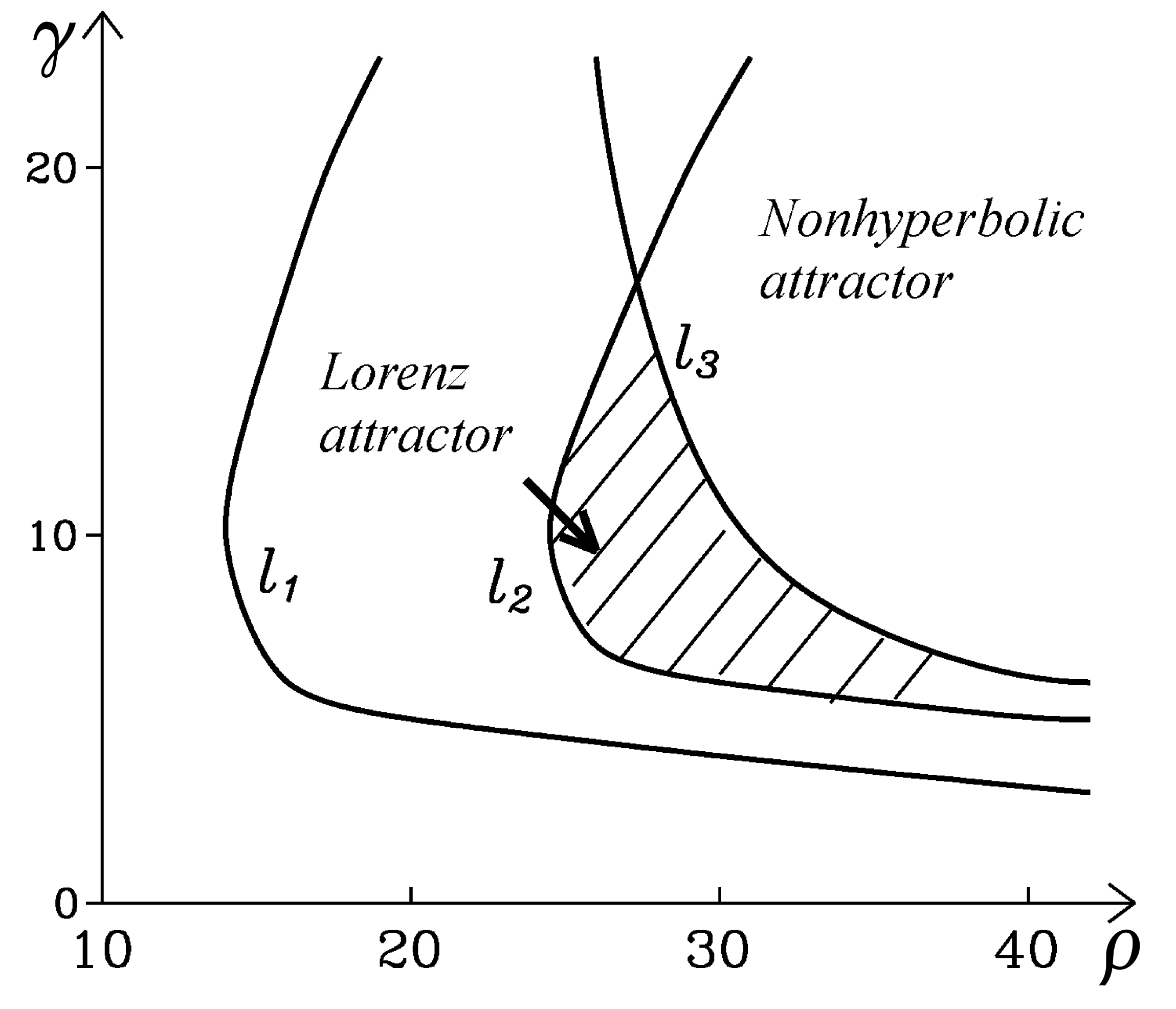}
\caption[]{Bifurcation diagram of the Lorenz system Eq.(\ref{eq:lorenz}) showing the hyperbolic regime of the Lorenz attractor (hatched) and the nonhyperbolic attractor for $b=8/3$.}
\label{fig7}
\end{figure}

Fig.~\ref{fig7} illustrates the bifurcation diagram of the Lorenz system (\ref{eq:lorenz}) in the $(\rho,\gamma)$ parameter plane for fixed parameter $b=8/3$ \cite{Bykov,ANI14}. The hatched area in the ($\rho,\gamma$) parameter space shows the hyperbolic regime where the Lorenz attractor is found. If $\gamma=10$ is fixed, then a transition to the regime of the nonhyperbolic attractor of the saddle-focus type occurs when the bifurcation line $l_3$ is crossed from left to right.

\begin{figure}[htbp]
\center{\includegraphics[width=.7\linewidth]{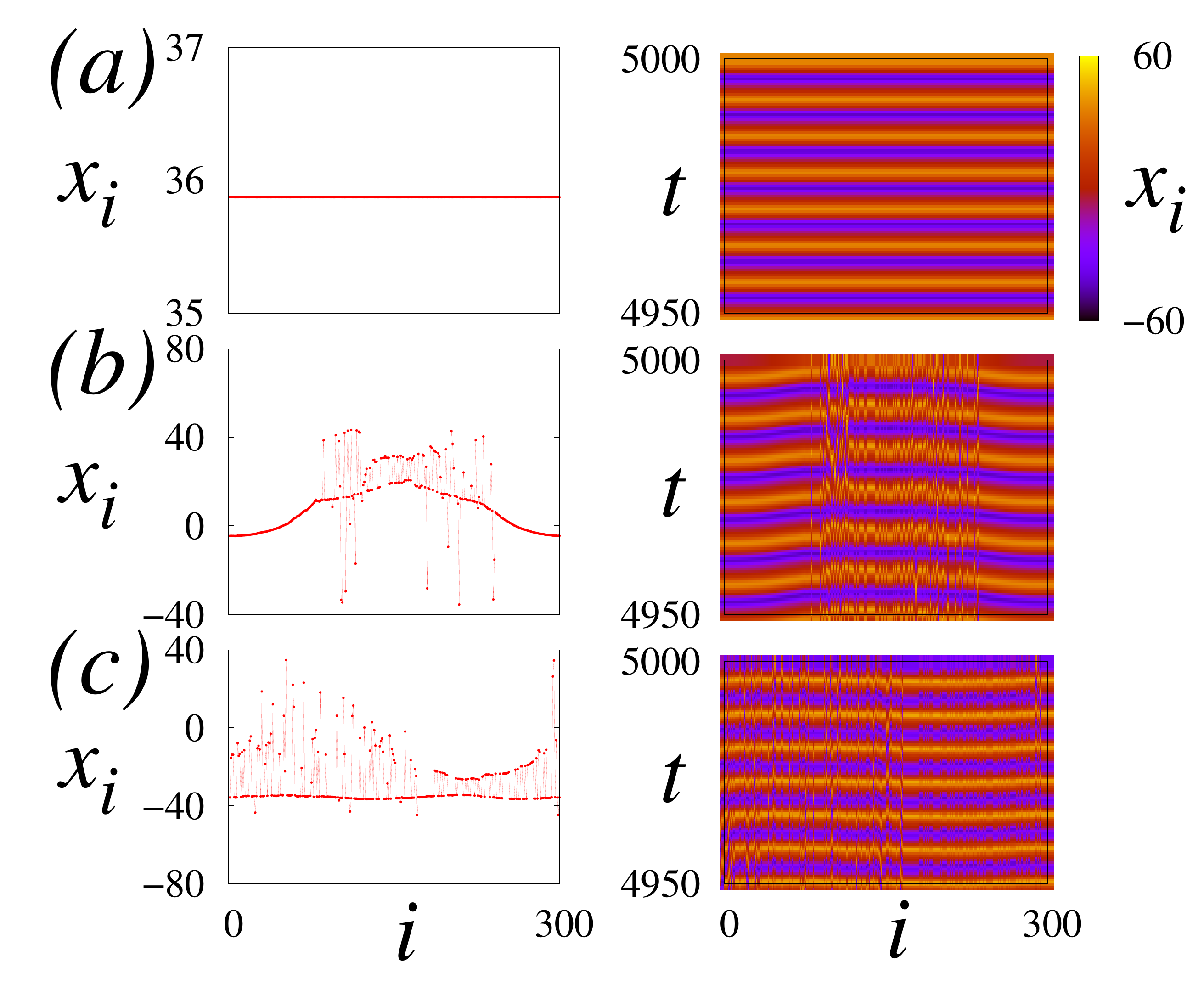}}
\caption[]{Chimera state in a ring of $N=300$ nonlocally coupled Lorenz systems in the nonhyperbolic regime. Snapshots at time $t=5000$ (left column) and space-time plots (right column) are shown. Parameters: $b=8/3$, $\gamma=10$, $\rho=220$, $r=0.22$. Decreasing from top to bottom, $\sigma=0.4$, $\sigma=0.35$ and $\sigma=0.1$. (a) -- complete synchronization, (b) -- chimera state, (c) -- spatial incoherence.}
\label{fig8}
\end{figure}

We simulate a ring of nonlocally coupled Lorenz systems \cite{LOR63} with parameters $\rho=220$, $\gamma=10$, $b=8/3$, corresponding
 to the nonhyperbolic regime. Indeed, we find a coherence-incoherence scenario via chimera states composed of one incoherent and one coherent domain as shown in Fig.~\ref{fig8}. The reason why the chimera does not consist of two incoherent and two coherent domains is
that it bifurcates from the $k=0$ (homogeneous) coherent profile, and not from the $k=1$ coherent profile as in Fig.4.  Such different types of chimeras have also been noted in the Van der Pol system \cite{OME15a}.

\section{Conclusions}
We have hypothesized that chimera states can be obtained in networks of chaotic nonhyperbolic systems and that they cannot be found in the networks of robust hyperbolic systems. This hypothesis has been confirmed by numerical simulations of a ring of nonlocally coupled time-discrete and time-continuous systems both for hyperbolic and nonhyperbolic cases.

Our results may be applied to a variety of diverse systems.
On one hand, chaotic behaviour can be found in numerous natural systems such as weather and climate, and chaotic dynamics plays an important role in many fields of science and engineering ranging from physics, chemistry, biology to economics and sociology.
On the other hand, there has been growing interest in chimera patterns with a wide scope of applications. Therefore, disclosing the universal mechanism of the emergence of chimeras in chaotic systems is a vital step in understanding the functionality of such complex systems.

\section{Acknowledgement}
This work was supported by DFG in the framework of SFB 910 and by the Russian Foundation for Basic Research (grant No.15-02-02288).

\section{Appendix: Critical coupling strength}\label{sub:strength}
In the following, we analyze the bifurcation scenario from complete coherence to complete incoherence via chimera states, and derive an analytic estimate of the critical coupling strength $\sigma_c$ for the onset of chimera states for the H\'enon and the Lozi map, respectively.

From a geometrical point of view and in the thermodynamic limit $N\rightarrow \infty$, where $r=P/N$ is fixed, coherent solutions $x_i^t$ approach a smooth profile  $x^t(\xi)$, where $\xi$ is the spatial variable, and the spatially continuous version of Eq.~(\ref{eq:coupling}) is given by
\begin{equation}\label{eq:continuous}
\begin{split}
x^{t+1}(\xi) = f(x^t(\xi),y^t(\xi)) + \frac{\sigma}{2r}\int_{\xi-r}^{\xi+r} \big[f\big(x^{t}(\eta), \\
y^{t}(\eta)\big) - f\left(x^{t}(\xi),x^{t}(\xi)\right)\big] d\eta; \qquad y^{t+1}(\xi) = \beta x^t(\xi).
\end{split}
\end{equation}
A transition from coherence to partial incoherence (chimera state) occurs if the respective profile $x^t(\xi), y^t(\xi)$ 
becomes discontinuous at some points $\xi$ of the ring $\mathcal{S}^{1}$.

Let us consider a solution of system~(\ref{eq:continuous}) with wave number $k=1$ (i.e., wavelength equals system size) and period-2 dynamics in time. Hence we can reduce the dynamics by even and odd time steps $x^0(\xi),y^0(\xi)$ and $x^1(\xi),y^1(\xi)$, respectively. This leads to $x^{1-j}(\xi) = (1-\sigma)f(x^j(\xi), y^j(\xi)) + \frac{\sigma}{2r} \int_{\xi-r}^{\xi+r} f\left(x^{j}(\eta),y^{j}(\eta)\right)  d\eta $; $y^{1-j}(\xi) = \beta x^j(\xi)$ 
with $j=0, 1$. Taking the spatial derivative $\partial_\xi x^j(\xi) \equiv x_{\xi}^j$ yields
\begin{eqnarray}
\label{eq:deriv}
x_{\xi}^{1-j}(\xi) = (1-\sigma)\left[f_{x^j} (x^j ,y^j) x_{\xi}^j + f_{y^j} (x^j,y^j) y_{\xi}^j\right] + \nonumber\\
+\frac{\sigma}{2r}\left[f\left(x^j(\xi+r),y^j(\xi+r)\right) - f\left(x^j(\xi-r),y^j(\xi-r)\right)\right]  \nonumber \\
y_{\xi}^{j}(\xi) = \beta x_{\xi}^{1-j}(\xi).
\end{eqnarray}
where $f_x(x,y)\equiv \partial_x f(x,y)$ etc.
At the point $\xi$ where the smooth profile breaks up, the spatial derivative becomes infinite. Considering that $|x_{\xi}^j|, |y_{\xi}^j|$ diverge to infinity, we can neglect the coupling term on the right-hand side of Eq.~(\ref{eq:deriv}). The main contribution comes from the first term. Using $f_y(x,y)=1$ we obtain:
\begin{align}\label{eq:deriv_}
x_{\xi}^{1-j}(\xi) =& (1-\sigma)\left[f_{x^j} (x^j ,y^j) x_{\xi}^j + \beta x_{\xi}^{1-j}\right] 
\end{align} 
Multiplying the equations for even and odd time steps ($j=0,1$) we obtain $x_{\xi}^{0}x_{\xi}^{1} = (1-\sigma)^2\left[f_{x^1} (x^1 ,y^1) x_{\xi}^1 + \beta x_{\xi}^{0}\right]\left[f_{x^0} (x^0 ,y^0) x_{\xi}^0 + \beta x_{\xi}^{1}\right]$.
Now assume the symmetry of the spatial profiles at the turning points (where the slope $m$ becomes infinite) $x_{\xi}^{1}=-x_{\xi}^{0}\equiv m$,
which yields the following condition
\begin{equation}\label{eq:cond}
1 = (1-\sigma)^2 \left[f_{x^1} (x^1 ,y^1) - \beta \right] \left[f_{x^0} (x^0 ,y^0) - \beta \right].
\end{equation} 
In order to obtain an analytic approximation for the critical coupling strength $\sigma_c$ at the onset of the chimera states,
we assume that the points $\xi$ where the smooth profile breaks up and the spatial derivative becomes infinite are fixed points 
($x^*, y^*$) of the map.
This yields $x^0=x^1=x^*$, $y^0=y^1=y^*$ and Eq.(\ref{eq:cond})  becomes
$1 = (1-\sigma)^2 \left[f_x (x^* ,y^*) - \beta \right]^2$. This leads to
\begin{equation}\label{eq:sigma_crit}
\sigma_c = 1 \pm \frac{1}{|f_x (x^* ,y^*)-\beta|}.
\end{equation} 
where the minus sign should be chosen since the lower value of $\sigma_c$ represents the threshold where the smooth profile breaks up with decreasing coupling strength $\sigma$, and the spatial coherence is lost.\\

For the Henon map $f(x,y)=1-\alpha x^2+y$ the derivative is equal to $f_x=-2\alpha x^*$, and we obtain $\sigma_c = 1 - \frac{1}{|2 \alpha x^*+\beta|}$.
The fixed points of the Henon map for $\alpha=1.4$, $\beta=0.3$ are approximately $x^*=0.63$ and $x^*=-1.13$.
Using $x^*=0.63$ we find $\sigma_c \approx 0.52$.

For the Lozi map $f(x,y)=1-\alpha |x|+y$ the derivative is equal to $f_x=-\alpha$ for $x^*>0$, and $f_x=\alpha$ for $x^*<0$,
and we obtain for $x^*<0$: $\sigma_c = 1 - \frac{1}{|\alpha -\beta|}$.
The fixed points of the Lozi map for $\alpha=1.4$, $\beta=0.3$ are approximately $x^*=0.48$ and $x^*=-1.43$.
Using $x^*<0$ we find $\sigma_c \approx 0.09$, which is so small that no period-2 dynamics exists,
since the dynamics is chaotic (similar to the uncoupled maps). Hence the above argument is not valid, and a transition to partial incoherence (chimera state) of the above type does not exist.\\

From the numerical simulations for the Henon map we obtain that the transition from coherence to incoherence in system~(\ref{eq:coupling}) occurs for a coupling strength $\sigma$ close to $0.5$.  The deviation from our approximation is due to
the finite number of nodes in the simulations, while the analytics assumes the continuum limit. 

\bibliography{ref}  

\begin{thebibliography}{10}
\expandafter\ifx\csname url\endcsname\relax\def\url#1{\texttt{#1}}\fi

\bibitem{PAN15}
\Name{Panaggio M.~J. \and Abrams D.~M.} \REVIEW{Nonlinearity}{28}{2015}{R67}.

\bibitem{KUR02a}
\Name{Kuramoto Y. \and Battogtokh D.} \REVIEW{Nonlin. Phen. in Complex
  Sys.}{5}{2002}{380}.

\bibitem{ABR04}
\Name{Abrams D.~M. \and Strogatz S.~H.}
  \REVIEW{Phys.~Rev.~Lett.}{93}{2004}{174102}.

\bibitem{MOT10}
\Name{Motter A.~E.} \REVIEW{Nature Physics}{6}{2010}{164}.

\bibitem{OME11}
\Name{Omelchenko I., Maistrenko Y., H{\"o}vel P. \and Sch{\"o}ll E.}
  \REVIEW{Phys. Rev. Lett.}{106}{2011}{}.

\bibitem{OME12}
\Name{Omelchenko I., Riemenschneider B., H{\"o}vel P., Maistrenko Y. \and
  Sch{\"o}ll E.} \REVIEW{Phys. Rev. E}{85}{2012}{}.

\bibitem{SET13}
\Name{Sethia G.~C., Sen A. \and Johnston G.~L.} \REVIEW{Phys. Rev.
  E}{88}{2013}{042917}.

\bibitem{SET14}
\Name{Sethia G.~C. \and Sen A.} \REVIEW{Phys. Rev. Lett.}{112}{2014}{144101}.

\bibitem{ZAK14}
\Name{Zakharova A., Kapeller M. \and Sch{\"o}ll E.}
  \REVIEW{Phys.~Rev.~Lett.}{112}{2014}{154101}.

\bibitem{ZAK15b}
\Name{Zakharova A., Kapeller M. \and Sch{\"o}ll E.} \REVIEW{J. Phys. Conf.
  Series, arXiv}{}{2015}{} 1503.03371.

\bibitem{BOE15}
\Name{B{\"o}hm F., Zakharova A., Sch{\"o}ll E. \and L{\"u}dge K.} \REVIEW{Phys.
  Rev. E}{91}{2015}{040901}.

\bibitem{OME15a}
\Name{Omelchenko I., Zakharova A., H{\"o}vel P., Siebert J. \and Sch{\"o}ll E.}
  \REVIEW{arXiv}{}{2015}{} 1503.03377.

\bibitem{BAS15}
\Name{Bastidas V., Omelchenko I., Zakharova A., Sch{\"o}ll E. \and Brandes T.}
  \REVIEW{arXiv}{}{2015}{} 1505.02639v1.

\bibitem{OME13}
\Name{Omelchenko I., Omel'chenko O.~E., H{\"o}vel P. \and Sch{\"o}ll E.}
  \REVIEW{Phys. Rev. Lett.}{110}{2013}{224101}.

\bibitem{OME15}
\Name{Omelchenko I., Provata A., Hizanidis J., Sch{\"o}ll E. \and H{\"o}vel P.}
  \REVIEW{Phys. Rev. E}{91}{2015}{022917}.

\bibitem{HIZ15}
\Name{Hizanidis J., Panagakou E., Omelchenko I., Sch{\"o}ll E., H{\"o}vel P.
  \and Provata A.} \REVIEW{arXiv}{}{2015}{} 1504.08125v1.

\bibitem{ROS14a}
\Name{Rosin D.~P., Rontani D., Haynes N.~D., Sch{\"o}ll E. \and Gauthier D.~J.}
  \REVIEW{Phys.~Rev.~ E}{90}{2014}{030902(R)}.

\bibitem{HAG12}
\Name{Hagerstrom A.~M., Murphy T.~E., Roy R., H{\"o}vel P., Omelchenko I. \and
  Sch{\"o}ll E.} \REVIEW{Nature Physics}{8}{2012}{658}.

\bibitem{TIN12}
\Name{Tinsley M.~R., Nkomo S. \and Showalter K.} \REVIEW{Nature
  Physics}{8}{2012}{662}.

\bibitem{MAR13}
\Name{Martens E.~A., Thutupalli S., Fourri{\`e}re A. \and Hallatschek O.}
  \REVIEW{Proc. Nat. Acad. Sciences}{110}{2013}{10563}.

\bibitem{KAP14}
\Name{Kapitaniak T., Kuzma P., Wojewoda J., Czolczynski K. \and Maistrenko Y.}
  \REVIEW{Scientific Reports}{4}{2014}{6379}.

\bibitem{LAR13}
\Name{Larger L., Penkovsky B. \and Maistrenko Y.} \REVIEW{Phys. Rev.
  Lett.}{111}{2013}{054103}.

\bibitem{GAM14}
\Name{Gambuzza L.~V., Buscarino A., Chessari S., Fortuna L., Meucci R. \and
  Frasca M.} \REVIEW{Phys. Rev. E}{90}{2014}{032905}.

\bibitem{LAR15}
\Name{Larger L., Penkovsky B. \and Maistrenko Y.}
  \REVIEW{arXiv}{1411.4483}{2015}{} accepted by Nature Comm.

\bibitem{WIC13}
\Name{Wickramasinghe M. \and Kiss I.~Z.} \REVIEW{PLoS ONE}{8}{2013}{e80586}.

\bibitem{SCH14a}
\Name{Schmidt L., Sch{\"o}nleber K., Krischer K. \and Garcia-Morales V.}
  \REVIEW{Chaos}{24}{2014}{013102}.

\bibitem{RAT00}
\Name{Rattenborg N.~C., Amlaner C.~J. \and Lima S.~L.} \REVIEW{Neurosci.
  Biobehav. Rev.}{24}{2000}{817}.

\bibitem{LAI01}
\Name{Laing C.~R. \and Chow C.~C.} \REVIEW{Neural Computation}{13}{2001}{1473}.

\bibitem{SAK06a}
\Name{Sakaguchi H.} \REVIEW{Phys. Rev.~E}{73}{2006}{031907}.

\bibitem{ROT14}
\Name{Rothkegel A. \and Lehnertz K.} \REVIEW{New J. of
  Phys.}{16}{2014}{055006}.

\bibitem{MOT13a}
\Name{Motter A.~E., Myers S.~A., Anghel M. \and Nishikawa T.} \REVIEW{Nature
  Physics}{9}{2013}{191}.

\bibitem{GON14}
\Name{Gonzalez-Avella J.~C., Cosenza M.~G. \and Miguel M.~S.}
  \REVIEW{Physica~A}{399}{2014}{24}.

\bibitem{DZI13}
\Name{Dziubak V., Maistrenko Y. \and Sch{\"o}ll E.} \REVIEW{Phys.
  Rev.~E}{87}{2013}{032907}.

\bibitem{HEN76}
\Name{H{\'e}non M.} \REVIEW{Commun. Math. Phys.}{50}{1976}{69}.

\bibitem{Sharkovskii}
\Name{Sharkovskii A.} \REVIEW{Int. J. Bifurcation Chaos}{5}{1995}{1263}.

\bibitem{ANI95}
\Name{Anishchenko V.~S.} \Book{{Dynamical Chaos - Models and Experiments}}
  (World Scientific, Singapore) 1995.

\bibitem{ANI14}
\Name{Anishchenko V.~S., Vadivasova T. \and Strelkova G.} \Book{{Deterministic
  Nonlinear Systems}} {Springer Series in Synergetics} (Springer) 2014.

\bibitem{Lozi}
\Name{Lozi R.} \REVIEW{Le Journal de Physique Colloques}{39}{1978}{C5}.

\bibitem{Bykov}
\Name{Bykov V. \and Shil{\'n}ikov A.} \REVIEW{Selecta Math.
  Sovietica}{}{1992}{376}.

\bibitem{LOR63}
\Name{Lorenz E.~N.} \REVIEW{J. ~Atmos. ~Sci.}{20}{1963}{130}.

\end{thebibliography}
\bibliographystyle{eplbib}

\end{document}